\documentclass[twocolumn,english,aps, pra, showpacs, showkeys]{revtex4}
\usepackage[T1]{fontenc}
\usepackage[latin9]{inputenc}
\usepackage{babel}

\usepackage{prettyref}
\usepackage{amsthm}
\usepackage{amsmath}
\usepackage[unicode=true, 
 bookmarks=true,bookmarksnumbered=false,bookmarksopen=false,
 breaklinks=false,pdfborder={0 0 0},backref=false,colorlinks=false]
 {hyperref}
\hypersetup{pdftitle={Geometric measure of entanglement compared to measures based on fidelity},
 pdfauthor={Alexander Streltsov},
 pdfkeywords={Keywords: entanglement measures, quantum information, quantum computation}}
 
\makeatletter
\@ifundefined{textcolor}{}
{%
 \definecolor{BLACK}{gray}{0}
 \definecolor{WHITE}{gray}{1}
 \definecolor{RED}{rgb}{1,0,0}
 \definecolor{GREEN}{rgb}{0,1,0}
 \definecolor{BLUE}{rgb}{0,0,1}
 \definecolor{CYAN}{cmyk}{1,0,0,0}
 \definecolor{MAGENTA}{cmyk}{0,1,0,0}
 \definecolor{YELLOW}{cmyk}{0,0,1,0}
 }
\theoremstyle{plain}
\newtheorem{thm}{Theorem}
  \theoremstyle{plain}
  \newtheorem{prop}[thm]{Proposition}

\usepackage{lmodern}
\newcommand{\bra}[1]{\langle #1|}
\newcommand{\ket}[1]{|#1\rangle}
\newcommand{\braket}[1]{\langle#1\rangle}

\makeatother

\begin{document}

\title{Geometric measure of entanglement compared to measures based on fidelity}

\author{Alexander Streltsov}

\email{alexander.streltsov@physik.uni-wuerzburg.de}

\affiliation{Universit\"{a}t W\"{u}rzburg, Institut f\"{u}r Theoretische Physik
und Astrophysik, 97074 W\"{u}rzburg, Germany}

\pacs{03.67.-a, 03.67.Mn}

\keywords{quantum information, entanglement measures}
\begin{abstract}
One of the biggest problems in the theory of quantum information is
the quantification of amount of entanglement in an arbitrary multipartite
mixed state. Different axiomatic and operational measures were proposed
so far. In this work we will establish a connection between geometric
measure of entanglement proposed in {[}Phys. Rev. A \textbf{68}, 042307
(2003){]} and measures based on fidelity. One result will be, that
geometric and revised geometric measure of entanglement proposed in
{[}J. Phys. A: Math. Theor. \textbf{40}, 3507 (2007){]} are equal.
Further a useful expression for fidelity is derived.
\end{abstract}
\maketitle

\section{Introduction}

Entanglement as a purely quantum mechanical property was already recognized
around 1935 by Einstein, Podolsky and Rosen \cite{Einstein}. 

A pure state of a bipartite quantum system is entangled if and only
if it can not be written as a product state. Consider the two qubit
state \[
\ket{\psi}=\frac{1}{\sqrt{2}}\left(\ket{\uparrow\downarrow}-\ket{\downarrow\uparrow}\right),\]
also called singlet state. One can show that this state can not be
written in the form $\ket{\psi}=\ket{a}\ket{b}$ with single qubit
states $\ket{a}$ and $\ket{b}$.

In this paper we consider the most general case, a multipartite mixed
state $\rho$ on a Hilbert space $\mathcal{H}=\otimes_{j=1}^{n}\mathcal{H}_{j}$.
It is in general called entangled if it can \emph{not} be written
in the form \begin{equation}
\rho=\sum_{i}p_{i}\left(\otimes_{j=1}^{n}\ket{\psi_{i}^{\left(j\right)}}\bra{\psi_{i}^{\left(j\right)}}\right),\label{eq:rho}\end{equation}
with non-negative probabilities $p_{i}$, $\sum_{i}p_{i}=1$, and
$\ket{\psi_{i}^{\left(j\right)}}$ being states on $\mathcal{H}_{j}$
\cite{Horodecki2009,Werner1989}. Otherwise the state is called separable.

Entanglement of pure bipartite states $\ket{\psi}$ is usually quantified
by the entanglement entropy \begin{eqnarray*}
E\left(\ket{\psi}\right) & = & -Tr\left[\rho^{A}\log_{2}\rho^{A}\right],\\
\rho^{A} & = & Tr_{B}\left[\ket{\psi}\bra{\psi}\right].\end{eqnarray*}
For mixed states different measures were proposed. In this paper we
will consider the \emph{geometric measure of entanglement} $E_{Ge}$
proposed in \cite{Wei2003} and compare it to measures based on fidelity
\cite{Biham2002,Cao2007,Shapira2006,Shimony1995,Vedral1998}. Those
are measures of the form \[
E_{f}\left(\rho\right)=f\left(\max_{\sigma\in S}F\left(\rho,\sigma\right)\right),\]
where $F$ is the quantum fidelity, $f$ is a proper chosen function
and $S$ is the set of separable states of the form \prettyref{eq:rho}.

By construction, $E_{Ge}$ was supposed to be different from measures
based on fidelity. In this paper we show that this is \emph{not} the
case. One of the main results of this paper will be, that $E_{Ge}$
is also a fidelity-based measure.

The structure of the paper is as follows. In \prettyref{sec:Definitions}
we give important definitions. In \prettyref{sec:Results} we give
main results for pure states, mixed states and two qubit states. A
conclusion is given in \prettyref{sec:Concluding-remarks}.

Appendix \ref{sec:Bipartite-pure-states} concentrates on bipartite
pure states and in Appendix \ref{sec:Optimal-purifications} we prove
a proposition needed for our main result.

\section{\label{sec:Definitions}Definitions}

First we restate the definition of fidelity $F\left(\rho,\sigma\right)$
between two quantum states $\rho$ and $\sigma$: \begin{equation}
F\left(\rho,\sigma\right)=\left(Tr\left[\sqrt{\sqrt{\rho}\sigma\sqrt{\rho}}\right]\right)^{2}.\end{equation}
It is important to notice that many authors define the fidelity as
the square root of our definition, important example are the authors
of \cite{Nielsen2000}.

In the following we always consider $n$-partite states on finite
dimensional Hilbert space $\mathcal{H}=\otimes_{i=1}^{n}\mathcal{H}_{i}$.

For a mixed state $\rho$ we now define the \emph{fidelity of separability}:
\begin{equation}
F_{sep}\left(\rho\right)=\max_{\sigma\in S}F\left(\rho,\sigma\right),\label{eq:Fsep}\end{equation}
maximization is done over the set of $n$-partite separable states
$S$ of the form $\sigma=\sum_{i}p_{i}\left(\otimes_{j=1}^{n}\ket{\psi_{i}^{\left(j\right)}}\bra{\psi_{i}^{\left(j\right)}}\right)$.

For pure states \emph{geometric measure of entanglement (GME)} is
defined as \cite{Shimony1995,Barnum2001} \begin{eqnarray}
E_{Ge}\left(\ket{\psi}\right) & = & 1-\Lambda_{max}^{2}\left(\ket{\psi}\right),\label{eq:GME}\\
\Lambda_{max}\left(\ket{\psi}\right) & = & \max_{\ket{\phi}\in S}\left|\braket{\phi|\psi}\right|,\label{eq:L}\end{eqnarray}
the maximization runs over all separable pure states $\ket{\phi}=\otimes_{i=1}^{n}\ket{\phi^{\left(i\right)}}$.
Extension to mixed states is made over the convex roof construction
\cite{Wei2003}: \begin{equation}
E_{Ge}\left(\rho\right)=\min_{\rho=\sum_{i}p_{i}\ket{\psi_{i}}\bra{\psi_{i}}}\sum_{i}p_{i}E_{Ge}\left(\ket{\psi_{i}}\right),\label{eq:GME-1}\end{equation}
minimization is done over all pure state decompositions of $\rho$. 

In \cite{Cao2007} authors defined the \emph{revised geometric measure
of entanglement (RGME)} \begin{equation}
E_{RGe}\left(\rho\right)=1-F_{sep}\left(\rho\right).\label{eq:RGE}\end{equation}

\emph{Groverian measure of entanglement} for a mixed state $\rho$
is defined as \cite{Biham2002,Shapira2006} \begin{equation}
E_{Gr}\left(\rho\right)=\sqrt{1-F_{sep}\left(\rho\right)}.\label{eq:Gr}\end{equation}

Finally \emph{Bures measure of entanglement} is defined as \cite{Vedral1998}
\begin{equation}
E_{B}\left(\rho\right)=2\left(1-\sqrt{F_{sep}\left(\rho\right)}\right).\label{eq:B}\end{equation}
As the measures \prettyref{eq:RGE}, \prettyref{eq:Gr} and \prettyref{eq:B}
are all simple functions of $F_{sep}$, we will only give results
for $F_{sep}$ in the following sections.

\section{\label{sec:Results}Results}

\subsection{Pure states}

In the following we will consider pure states $\ket{\psi}\in\otimes_{i=1}^{n}\mathcal{H}_{i}$.
\begin{prop}
\label{pro:Fsep}For pure state $\ket{\psi}\in\otimes_{i=1}^{n}\mathcal{H}_{i}$
holds: \begin{equation}
F_{sep}\left(\ket{\psi}\right)=\max_{\ket{\phi}\in S}\left|\braket{\phi|\psi}\right|^{2}=\Lambda_{max}^{2}\left(\ket{\psi}\right),\label{eq:Fsep-1}\end{equation}
maximization is done over separable pure states $\ket{\phi}=\otimes_{j=1}^{n}\ket{\phi^{\left(j\right)}}$.\end{prop}
\begin{proof}
To evaluate $F_{sep}$ for a pure state $\ket{\psi}$ we need to find
a separable state $\sigma$ that maximizes the fidelity among all
separable states, such that \begin{equation}
F\left(\ket{\psi},\sigma\right)=\max_{\rho_{sep}\in S}F\left(\ket{\psi},\rho_{sep}\right).\label{eq:F}\end{equation}
Then $F_{sep}\left(\ket{\psi}\right)=F\left(\ket{\psi},\sigma\right)$.
Set now $\sigma=\sum_{i}q_{i}\ket{\phi_{i}}\bra{\phi_{i}}$ with separable
states $\ket{\phi_{i}}=\otimes_{j=1}^{n}\ket{\phi_{i}^{\left(j\right)}}$.
Then we see:\begin{equation}
F\left(\ket{\psi},\sigma\right)=\braket{\psi|\sigma|\psi}=\sum_{i}q_{i}\left|\braket{\psi|\phi_{i}}\right|^{2}.\label{eq:F-1}\end{equation}
We define $\ket{\phi_{1}}$ to have the largest overlap with $\ket{\psi}$,
that is $\left|\braket{\psi|\phi_{1}}\right|\geq\left|\braket{\psi|\phi_{i}}\right|$.
From \prettyref{eq:F-1} follows: \begin{equation}
F\left(\ket{\psi},\sigma\right)\leq\sum_{i}q_{i}\left|\braket{\psi|\phi_{1}}\right|^{2}=\left|\braket{\psi|\phi_{1}}\right|^{2}=F\left(\ket{\psi},\ket{\phi_{1}}\right).\end{equation}
In maximization \prettyref{eq:F} we can restrict ourselves to pure
states, such that $\sigma$ can be chosen to be pure: $\sigma=\ket{\phi}\bra{\phi}$,
$\ket{\phi}=\otimes_{i=1}^{n}\ket{\phi^{\left(i\right)}}$. As $F\left(\ket{\psi},\ket{\phi}\right)=\left|\braket{\phi|\psi}\right|^{2}$
the proof is complete.
\end{proof}
In \cite{Shimony1995} the author showed that for bipartite pure states
$\ket{\psi}$ with Schmidt decomposition $\ket{\psi}=\sum_{i}\lambda_{i}\ket{i^{\left(1\right)}}\otimes\ket{i^{\left(2\right)}}$
the overlap $\Lambda_{max}\left(\ket{\psi}\right)$ is given by the
largest Schmidt coefficient \begin{equation}
\Lambda_{max}\left(\ket{\psi}\right)=\lambda=\max_{i}\left\{ \lambda_{i}\right\} .\label{eq:l}\end{equation}
An alternative proof can be found in Appendix \ref{sec:Bipartite-pure-states}.
Thus we can state the following proposition.
\begin{prop}
For bipartite pure state $\ket{\psi}$ with largest Schmidt coefficient
$\lambda$ holds: \begin{eqnarray}
F_{sep}\left(\ket{\psi}\right) & = & \lambda^{2}.\end{eqnarray}

\end{prop}

\subsection{Mixed states}

Now we consider mixed states $\rho$ on a finite dimensional Hilbert
space $\mathcal{H}=\otimes_{i=1}^{n}\mathcal{H}_{i}$. A purification
of $\rho$ is a pure state $\ket{\psi}\in\mathcal{H}_{0}\otimes\mathcal{H}$
such that $\rho=Tr_{0}\left[\ket{\psi}\bra{\psi}\right]$.
\begin{prop}
\label{pro:Purifications}Let $\rho$ be a mixed quantum state with
particular decomposition $\left\{ p_{i},\ket{\psi_{i}}\right\} $
such that $\rho=\sum_{i}p_{i}\ket{\psi_{i}}\bra{\psi_{i}}$. Every
purification of $\rho$ can be written in the form \begin{equation}
\ket{\psi}=\sum_{i}\sqrt{p_{i}}\ket{\psi_{i}^{\left(0\right)}}\ket{\psi_{i}}\label{eq:psi}\end{equation}
with $\ket{\psi_{i}^{\left(0\right)}}\in\mathcal{H}_{0}$, $\braket{\psi_{i}^{\left(0\right)}|\psi_{j}^{\left(0\right)}}=\delta_{ij}$.\end{prop}
\begin{proof}
Let $\ket{\phi}$ be an arbitrary purification of $\rho$ with Schmidt
decomposition \begin{equation}
\ket{\phi}=\sum_{i}\sqrt{q_{i}}\ket{\phi_{i}^{\left(0\right)}}\ket{\phi_{i}}.\label{eq:phi}\end{equation}
$\ket{\phi_{i}}$ are eigenstates and $q_{i}$ are corresponding eigenvalues
of $\rho$. According to \cite[Theorem 2.6 on page 103]{Nielsen2000}
there is a unitary matrix $u$ such that \begin{equation}
\sqrt{q_{i}}\ket{\phi_{i}}=\sum_{j}u_{ij}\sqrt{p_{j}}\ket{\psi_{j}}.\label{eq:ens}\end{equation}
With \prettyref{eq:ens} in \prettyref{eq:phi} we get \[
\ket{\phi}=\sum_{j}\sqrt{p_{j}}\ket{\psi_{j}^{\left(0\right)}}\ket{\psi_{j}}\]
with $\ket{\psi_{j}^{\left(0\right)}}=\sum_{i}u_{ij}\ket{\phi_{i}^{\left(0\right)}}$,
and thus $\braket{\psi_{i}^{\left(0\right)}|\psi_{j}^{\left(0\right)}}=\delta_{ij}$.
This is exactly the form \prettyref{eq:psi}, this ends the proof.
\end{proof}
A direct consequence of Proposition \prettyref{pro:Purifications}
is that every purification of a separable state of the form \prettyref{eq:rho}
can be written in the form \[
\ket{\psi}=\sum_{i}\sqrt{p_{i}}\left(\otimes_{j=0}^{n}\ket{\psi_{i}^{\left(j\right)}}\right)\]
with $\braket{\psi_{i}^{\left(0\right)}|\psi_{j}^{\left(0\right)}}=\delta_{ij}$,
$\sum_{i}p_{i}=1$, $p_{i}\geq0$.

With this in mind we can prove the following theorem.
\begin{thm}
\label{thm:Fsep}For a multipartite mixed state $\rho$ on a finite
dimensional Hilbert space $\mathcal{H}=\otimes_{j=1}^{n}\mathcal{H}_{j}$
holds: \begin{equation}
F_{sep}\left(\rho\right)=\max_{\rho=\sum_{i}p_{i}\ket{\psi_{i}}\bra{\psi_{i}}}\sum_{i}p_{i}F_{sep}\left(\ket{\psi_{i}}\right),\end{equation}
maximization is done over all pure state decompositions of $\rho$.\end{thm}
\begin{proof}
Let $\sigma$ be a separable state that maximizes the fidelity among
all separable states, that is $F_{sep}\left(\rho\right)=F\left(\rho,\sigma\right)$.
According to \cite[Equation 9.72 on page 411]{Nielsen2000} holds:
\begin{equation}
F\left(\rho,\sigma\right)=\max_{\sigma=Tr_{0}\left[\ket{\phi}\bra{\phi}\right]}\left|\braket{\psi|\phi}\right|^{2},\label{eq:F-2}\end{equation}
where $\ket{\psi}$ is a purification of $\rho$ and the maximization
is done over all purifications of $\sigma$. In the following $\ket{\phi}$
will denote a particular purification that realizes the maximum: $F\left(\rho,\sigma\right)=\left|\braket{\psi|\phi}\right|^{2}$.

Using Proposition \prettyref{pro:Purifications} we write \begin{equation}
\ket{\phi}=\sum_{i}\sqrt{q_{i}}\otimes_{j=0}^{n}\ket{\phi_{i}^{\left(j\right)}}\label{eq:ph}\end{equation}
with $\sum_{i}q_{i}=1$ and $\braket{\phi_{i}^{\left(0\right)}|\phi_{j}^{\left(0\right)}}=\delta_{ij}$.
We prove in Appendix \ref{sec:Optimal-purifications} that \begin{equation}
\left|\braket{\psi|\phi}\right|^{2}=\sum_{i}\max_{\braket{\phi_{i}^{\left(0\right)}|\phi_{j}^{\left(0\right)}}=\delta_{ij}}\left|\braket{\psi|\otimes_{j=0}^{n}\phi_{i}^{\left(j\right)}}\right|^{2}.\label{eq:psiphi}\end{equation}
Using Proposition \prettyref{pro:Purifications} we write: $\ket{\psi}=\sum_{i}\sqrt{p_{i}}\ket{\psi_{i}^{\left(0\right)}}\ket{\psi_{i}}$.
Noting that there always is a unitary matrix $u$ such that $\ket{\psi_{i}^{\left(0\right)}}=\sum_{j}u_{ij}\ket{\phi_{j}^{\left(0\right)}}$
we rewrite $\ket{\psi}$ as follows: \begin{eqnarray}
\ket{\psi} & = & \sum_{i,j}\sqrt{p_{i}}u_{ij}\ket{\phi_{j}^{\left(0\right)}}\ket{\psi_{i}}=\sum_{j}\sqrt{p_{j}'}\ket{\phi_{j}^{\left(0\right)}}\ket{\psi_{j}'},\label{eq:psi-1}\end{eqnarray}
where $\sqrt{p_{j}'}\ket{\psi_{j}'}=\sum_{i}u_{ij}\sqrt{p_{i}}\ket{\psi_{i}}$
and $\rho=\sum_{j}p_{j}'\ket{\psi_{j}'}\bra{\psi_{j}'}$. For simplicity
we write $p_{i}$ instead of $p_{i}'$ and $\ket{\psi_{i}}$ instead
of $\ket{\psi_{i}'}$.

With \prettyref{eq:psi-1} in \prettyref{eq:psiphi} we get: \begin{eqnarray}
\left|\braket{\psi|\phi}\right|^{2} & = & \max\sum_{i}p_{i}\left|\braket{\psi_{i}|\otimes_{j=1}^{n}\phi_{i}^{\left(j\right)}}\right|^{2},\end{eqnarray}
the maximum is taken over all pure state decompositions $\left\{ p_{i},\ket{\psi_{i}}\right\} $
such that $\rho=\sum_{i}p_{i}\ket{\psi_{i}}\bra{\psi_{i}}$ and over
all $\ket{\phi_{i}^{\left(j\right)}}\in\mathcal{H}_{j}$. With \prettyref{eq:Fsep-1}
we get  \begin{eqnarray*}
\left|\braket{\psi|\phi}\right|^{2} & = & \max_{\rho=\sum_{i}p_{i}\ket{\psi_{i}}\bra{\psi_{i}}}\sum_{i}p_{i}\Lambda_{max}^{2}\left(\ket{\psi_{i}}\right)\\
 & = & \max_{\rho=\sum_{i}p_{i}\ket{\psi_{i}}\bra{\psi_{i}}}\sum_{i}p_{i}F_{sep}\left(\ket{\psi_{i}}\right).\end{eqnarray*}
This ends the proof.
\end{proof}
Using \prettyref{thm:Fsep} we can prove the following proposition.
\begin{prop}
Geometric and revised geometric measure of entanglement are equal:
\begin{equation}
E_{Ge}\left(\rho\right)=1-F_{sep}\left(\rho\right).\label{eq:GME-2}\end{equation}
\end{prop}
\begin{proof}
Using definition \prettyref{eq:GME}, \prettyref{eq:GME-1} of $E_{Ge}$
and \prettyref{eq:Fsep-1} we write \begin{eqnarray}
E_{Ge}\left(\rho\right) & = & \min\sum_{i}p_{i}\left(1-F_{sep}\left(\ket{\psi_{i}}\right)\right),\end{eqnarray}
minimization is done over all pure state decompositions of $\rho$.
Using $\sum_{i}p_{i}=1$ this becomes \begin{equation}
E_{Ge}\left(\rho\right)=1-\max\sum_{i}p_{i}F_{sep}\left(\ket{\psi_{i}}\right).\end{equation}
Using \prettyref{thm:Fsep} the proof is complete.
\end{proof}

\subsection{Two qubits}

In \cite{Wei2003} geometric measure for two qubit states was derived:
\begin{equation}
E_{Ge}\left(\rho\right)=\frac{1}{2}\left(1-\sqrt{1-C\left(\rho\right)^{2}}\right)\label{eq:Ege-4}\end{equation}
with concurrence $C\left(\rho\right)$ \cite{Wootters1998}. With
\prettyref{eq:GME-2} we can compute $F_{sep}$ for two qubit states:
\begin{equation}
F_{sep}\left(\rho\right)=\frac{1}{2}\left(1+\sqrt{1-C\left(\rho\right)^{2}}\right).\label{eq:Fsep-2}\end{equation}
Using \prettyref{eq:Fsep-2} all entanglement measures based on fidelity
can be computed for two qubit states. For Bures measure of entanglement
we get \[
E_{B}\left(\rho\right)=2-2\sqrt{\frac{1+\sqrt{1-C\left(\rho\right)^{2}}}{2}},\]
we already presented this result in \cite{Streltsov2009}.

\section{\label{sec:Concluding-remarks}Concluding remarks}

In this paper we established a simple connection between the geometric
measure of entanglement and entanglement measures based on fidelity.
Using it, all results obtained for geometric measure can also be used
for other measures and vice versa.

One of our main results is \prettyref{thm:Fsep}. In words it can
be expressed as follows: the fidelity of separability is an upside
down version of a convex roof measure of entanglement. This result
underlines the importance of fidelity for quantum information theory,
especially for construction of entanglement measures. We believe that
it will be useful for further research in this direction.
\begin{acknowledgments}
I thank Christian Gogolin and Dagmar Bru{\ss} for discussion.
\end{acknowledgments}
\appendix

\section{\label{sec:Bipartite-pure-states}Bipartite pure states}

Let $\ket{\psi}$ be a bipartite pure state with Schmidt decomposition
\[
\ket{\psi}=\sum_{i}\lambda_{i}\ket{i^{\left(1\right)}}\otimes\ket{i^{\left(2\right)}}.\]
Further we define $\lambda=\underset{i}{\max}\left\{ \lambda_{i}\right\} $.
We will now show that $\lambda=\underset{\ket{\phi}\in S}{\max}\left|\braket{\psi|\phi}\right|$,
$\ket{\phi}=\ket{\phi^{\left(1\right)}}\otimes\ket{\phi^{\left(2\right)}}$.
Another proof was given in \cite{Shimony1995}.

First we rewrite the states as follows: \begin{eqnarray*}
\ket{\phi^{\left(1\right)}}=\sum_{i}a_{i}^{\star}\ket{i^{\left(1\right)}}, &  & \ket{\phi^{\left(2\right)}}=\sum_{i}b_{i}\ket{i^{\left(2\right)}}.\end{eqnarray*}
Now note that \begin{equation}
\left|\braket{\psi|\phi}\right|=\left|\sum_{i}\lambda_{i}a_{i}^{\star}b_{i}\right|=\left|\braket{a|Y|b}\right|\label{eq:Y}\end{equation}
with $Y$ being diagonal matrix with entries $\lambda_{i}$. $\ket{a}$
and $\ket{b}$ are normalized vectors with entries $a_{i}$ and $b_{i}$.

We have to maximize \prettyref{eq:Y} over all normalized vectors
$\ket{a}$ and $\ket{b}$. For this we will prove the following theorem:
\begin{thm}
\label{thm:Hermitian}For a Hermitian matrix $H$ with eigenvalues
$\lambda_{i}$ and two normalized vectors $\ket{a}$ and $\ket{b}$
holds:\begin{equation}
\left|\braket{a|H|b}\right|\leq\max_{i}\left|\lambda_{i}\right|.\label{eq:H}\end{equation}
\end{thm}
\begin{proof}
We will maximize $\left|\braket{a|H|b}\right|^{2}$: \[
\left|\braket{a|H|b}\right|^{2}=\braket{a|Z|a},\]
where $Z=H\ket{b}\bra{b}H$ is a Hermitian matrix with rank 1, thus
the only nonzero eigenvalue of $Z$ is $\braket{b|H^{2}|b}$. Further
from \cite[Theorem 4.2.2 on page 176]{Horn1985} follows that $\braket{a|Z|a}\leq\braket{b|H^{2}|b}\leq\underset{i}{\max}\lambda_{i}^{2}$.
This ends the proof.
\end{proof}
Using \prettyref{eq:H} in \prettyref{eq:Y} and noting that $Y\geq0$
we immediately get \[
\left|\braket{\psi|\phi}\right|\leq\lambda\]
for all separable states $\ket{\phi}$ and equality is attained if
$\ket{\phi}=\ket{1^{\left(1\right)}}\ket{1^{\left(2\right)}}$, where
$\lambda_{1}=\lambda$. This proves that \[
\underset{\ket{\phi}\in S}{\max}\left|\braket{\psi|\phi}\right|=\lambda,\]
as stated above.

\section{\label{sec:Optimal-purifications}Optimal purifications}

Let $\sigma$ be a separable state on $n$-partite Hilbert space $\mathcal{H}=\otimes_{i=1}^{n}\mathcal{H}_{i}$.
Then it can be written as \begin{equation}
\sigma=\sum_{i}q_{i}\left(\otimes_{j=1}^{n}\ket{\phi_{i}^{\left(j\right)}}\bra{\phi_{i}^{\left(j\right)}}\right).\end{equation}
According to Proposition \prettyref{pro:Purifications} every purification
of $\sigma$ can be written as a state $\ket{\phi}\in\mathcal{H}_{0}\otimes\mathcal{H}$
of the form \begin{equation}
\ket{\phi}=\sum_{i}\sqrt{q_{i}}\otimes_{j=0}^{n}\ket{\phi_{i}^{\left(j\right)}}\label{eq:phi-1}\end{equation}
with $\braket{\phi_{i}^{\left(0\right)}|\phi_{j}^{\left(0\right)}}=\delta_{ij}$,
$\sum_{i}q_{i}=1$.

For a given pure state $\ket{\psi}\in\mathcal{H}_{0}\otimes\mathcal{H}$
we now want to maximize $\left|\braket{\psi|\phi}\right|^{2}$ among
all states $\ket{\phi}$ of the form \prettyref{eq:phi-1}. Noting
that \begin{equation}
\left|\braket{\psi|\phi}\right|^{2}=\left|\sum_{i}\sqrt{q_{i}}\braket{\psi|\otimes_{j=0}^{n}\phi_{i}^{\left(j\right)}}\right|^{2}\label{eq:psiphi-1}\end{equation}
 we will now show that $q_{i}$ can be eliminated.
\begin{prop}
For any pure state $\ket{\psi}\in\mathcal{H}_{0}\otimes\mathcal{H}$
holds: \begin{equation}
\max_{Tr_{0}\left[\ket{\phi}\bra{\phi}\right]\in S}\left|\braket{\psi|\phi}\right|^{2}=\sum_{i}\max_{\braket{\phi_{i}^{\left(0\right)}|\phi_{j}^{\left(0\right)}}=\delta_{ij}}\left|\braket{\psi|\otimes_{j=0}^{n}\phi_{i}^{\left(j\right)}}\right|^{2}.\label{eq:max-1}\end{equation}
\end{prop}
\begin{proof}
Note that following inequality holds: \begin{equation}
\left|\braket{\psi|\phi}\right|=\left|\sum_{i}\sqrt{q_{i}}\braket{\psi|\otimes_{j=0}^{n}\phi_{i}^{\left(j\right)}}\right|\leq\sum_{i}\sqrt{q_{i}}\left|\braket{\psi|\otimes_{j=0}^{n}\phi_{i}^{\left(j\right)}}\right|.\label{eq:psiphi-2}\end{equation}
In maximizing $\left|\braket{\psi|\phi}\right|$ we are free to choose
the phases of $\ket{\phi_{i}^{\left(j\right)}}$, this can always
be done such that on rhs of \prettyref{eq:psiphi-2} equality holds,
that is \begin{equation}
\max_{Tr_{0}\left[\ket{\phi}\bra{\phi}\right]}\left|\braket{\psi|\phi}\right|=\max\sum_{i}\sqrt{q_{i}}\left|\braket{\psi|\otimes_{j=0}^{n}\phi_{i}^{\left(j\right)}}\right|.\label{eq:psiphi-3}\end{equation}
Maximization on rhs is done over all $\ket{\phi_{i}^{\left(j\right)}}$
with the only restriction $\braket{\phi_{i}^{\left(0\right)}|\phi_{j}^{\left(0\right)}}=\delta_{ij}$,
and over all $q_{i}$ restricted by $\sum_{i}q_{i}=1$. Maximization
over $q_{i}$ can be evaluated using Lagrange multipliers with the
result \begin{equation}
\sqrt{q_{i}}=\frac{\left|\braket{\psi|\otimes_{j=0}^{n}\phi_{i}^{\left(j\right)}}\right|}{\sqrt{\sum_{i}\left|\braket{\psi|\otimes_{j=0}^{n}\phi_{i}^{\left(j\right)}}\right|^{2}}}.\label{eq:q}\end{equation}
Using \prettyref{eq:q} in \prettyref{eq:psiphi-3} we get \prettyref{eq:max-1}.
This ends the proof.
\end{proof}
\begingroup\raggedright\endgroup

\end{document}